\documentclass[merge]{aastex63}  
\newcommand\Deltafdif{\Delta f_\mathrm{dif}}  
\newcommand\deltatcor{\delta t_\mathrm{cor}}
\newcommand\ds{d_\mathrm{s}}

\newcommand\Rb{R_I^\mathrm{b}}
\newcommand\Rc{R_I^\mathrm{c}}
\newcommand\rhodif{\rho_\mathrm{dif}}

\newcommand\tausc{\tau_\mathrm{sc}}    
\newcommand\tdif{t_\mathrm{dif}}       
\newcommand\thetares{\theta_\mathrm{res}}    
\newcommand\thetasc{\theta_\mathrm{sc}}    
\newcommand\tint{t_\mathrm{int}}
\newcommand\tref{t_\mathrm{ref}}       
\newcommand\Tscan{T_\mathrm{scan}}     
\newcommand\tscint{t_\mathrm{scint}}   
\newcommand\Tvis{T_\mathrm{vis}}
\received{08-26-2019}
\revised{11-21-2019}
\accepted{11-29-2019}
\journalinfo{Astrophysical~journal}

\shorttitle{Substructure  of visibility}
\shortauthors{Popov et al.}

\begin{document}

\title{Substructure of visibility functions from scattered radio emission of pulsars through space VLBI}

\correspondingauthor{M.~S.\ Burgin}
\email{mburgin@asc.rssi.ru}

\author[0000-0001-7931-646X]{M.~V.\ Popov}
\affiliation{Lebedev Physical Institute, Astro Space Center, Profsoyuznaya 84/32, Moscow, 117997 Russia}

\author{N.\ Bartel}
\affiliation{York University, 4700 Keele St., Toronto, ON M3J 1P3, Canada}

\author[0000-0002-0579-2938]{M.~S.\ Burgin}
\affiliation{Lebedev Physical Institute, Astro Space Center, Profsoyuznaya 84/32, Moscow, 117997 Russia}

\author[0000-0002-6936-4430]{C.~R.\ Gwinn}
\affiliation{University of California at Santa Barbara, Santa Barbara, CA 93106-4030, USA}

\author{T.~V.\ Smirnova}
\affiliation{Lebedev Physical Institute, Pushchino Radio Astronomy Observatory, Pushchino 142290, Moscow region, Russia}

\author{V.~A.\ Soglasnov}
\affiliation{Lebedev Physical Institute, Astro Space Center, Profsoyuznaya 84/32, Moscow, 117997 Russia}

\begin{abstract}
We report on the substructure of visibility functions in the delay domain of
PSRs B0329+54, B0823+26, B0834+06, B1933+16 and B0833-45 (Vela) observed with
earth-earth and RadioAstron space-earth two-element interferometers at
frequencies of 324 MHz and 
1668
MHz. All visibility functions display
unresolved spikes distributed over a range of delays. They are due to
band-limited scintillation noise and related to the scattering time. 
The envelopes for each but the
Vela pulsar are well fit by a single Lorentzian which we interpret as being
indicative of isotropic scattering on the plane of the sky due to a thin
scattering screen between the pulsar and us.  In contrast, the envelope for the
Vela pulsar needs to be mostly fit by at least two Lorentzians, a narrow and a
broad one at the same zero delay.  We interpret this characteristic as
indicative of 
  anisotropic scattering due to
more complex structure of scattering screens in the supernova remnant.  
  The possibility of describing the
  delay visibility functions by Lorentzians is likely a general property of
  pulsars and offers a new way of describing scattering parameters of the
  intervening interstellar medium.  Furthermore, for all our pulsars, the
  unresolved spikes in visibility functions of similar projected baselines were
  well correlated indicating that the telescopes are located in the same
  diffraction spot. The correlation vanished for visibilities from largely
  different baselines, when some radio telescopes are not in the same spot.
\end{abstract}


\section{Introduction}

Radio emission of compact celestial radio sources in our Galaxy and beyond can
be strongly scattered by inhomogeneities of the interstellar medium (ISM)
located between the source and the observer. This process causes angular
broadening of the source image, distortion of radio spectra, and intensity
fluctuations or scintillations of the radio emission (see, e.g.,
\cite{prokhorov1975,rickett1977,gwinn1998,shishov2003}).  Here we focus on the
effects of scattering on VLBI and space-VLBI observations of compact
sources at frequencies at which these effects are strong. We have chosen pulsars
as targets since they are intrinsically point-like even when observed with
space-VLBI on baselines as long as 200,000 km as provided by RadioAstron
  \citep{2013ARep...57..153K}.
Therefore the
structure of the source does not need to be considered in the analysis, and the
results are essentially exclusively due to scattering characteristics of the ISM
(see, e.g., \cite{gwinn2015, johnson2016, john_nar2016} for recent studies on
this subject).  In previous studies some characteristics of the scattering
screens in the ISM in terms of the pulsar's scintillation time, $\tscint$,
scattering time, $\tausc$, angular size of the scattering disk, 
$\thetasc$, 
and
decorrelation bandwidth, $\Deltafdif$, were already obtained. Assuming a single
thin scattering screen and combining, $\tausc$ with $\thetasc$, 
the distance,
$\ds$ of the scattering screen relative to the distance, $D$, of the pulsar
could be determined.  An analysis of these measurements indicates a possible
layered structure of the interstellar plasma in our Galaxy
\citep{fadeev2018,gwinn2016,popov2016,popov2017,popov2019}. 
In previous work, \citet{popov2016} 
presented an example of the VLBI visibility function for the pulsar PSR B1749-28.
They found for the first time that 
for this pulsar the dependence of the  average visibility function on delay
can be well fit by a Lorentzian.
In this paper we follow \citet{popov2016} and
focus on five more pulsars, 
four of them 
older pulsars and one of them 
the young Vela 
pulsar, PSR B0833-45, 
which is still embedded in its supernova remnant and also likely in the larger
Gum Nebula.  We selected the pulsars on the basis of their peak flux
  density in order to get a sufficiently high signal to noise ratio for the
  analysis, and on the basis of the selected pulsars having a relatively large
  range of dispersion measures.
  Further, the scattering time, $\tausc$,
  needed to be large enough so that a sufficiently large number of points of the
  envelope could be used for the model fit.  Since the bandwidth of our VLBI and
  space-VLBI observations was 16 MHz, the sampling step in delay of our
  visibility functions was 31.25 ns. That restricted our choice to pulsars with
  $\tausc \gtrsim 0.5 \mu$s so that at least a dozen of sampling points could
  be used for the fit. An additional concern was the selection of the
  observing frequency. Usually our first choice was to use data obtained at the
  lowest of the available frequencies, namely at 324 MHz. However, for two
  pulsars the dispersion measure was so high, that we needed to select the next
  higher available frequency, namely 1668 MHz to allow for a good fit of the
  visibility
  function.

Table~\ref{tab:param0} lists the pulsars with their periods, dispersion
measures, distances, galactic coordinates, observing frequency, 
 scattering parameters 
obtained at the observing frequencies 
 and the ratios
of the distance of the scattering screen relative to the distance of the pulsar,
obtained in our previous publications cited above. 
This is a small but
  somewhat representative list of pulsars with respect to the range of
  dispersion measures, galactic latitudes and scattering times.

\begin{deluxetable*}{lcccrrcclclcc}
\tablecaption{Parameters of pulsars\label{tab:param0}}
\tablehead{
 \colhead{PSR} & \colhead{$P$} & \colhead{$DM$} & \colhead{$D$}& \colhead{$l$} & \colhead{$b$} & \colhead{$\nu$} 
 & \colhead{$\tscint$}  & \colhead{$\tausc$} & \colhead{$\thetasc$} & \colhead{$\Deltafdif$} & \colhead{${\ds}/{D}$} & \colhead{Reference}
  \\
 &\colhead{(s)} & \colhead{(pc $\text{cm}^{-3}$)}  &\colhead{(kpc)} &\colhead{(deg)} &\colhead{(deg)}  &\colhead{(MHz)}
 &\colhead{(s)} &\colhead{($\mu$s)} &\colhead{(mas)} & \colhead{(kHz)}
 \\
  \colhead{(1)} & \colhead{(2)} & \colhead{(3)} & \colhead{(4)} & \colhead{(5)} & \colhead{(6)} &  \colhead{(7)}
  & \colhead{(8)} & \colhead{(9)} & \colhead{(10)} & \colhead{(11)} & \colhead{(12)} & \colhead{(13)}
} 
\startdata
    B0329+54 & 0.714 & \phn26.7  & 1.03 & 145.0 & -1.2 & \phn324  & 110                 & 4.1  & 4.8      & \phn\phn7.0 & 0.60      & 1 \\
    B0823+26 & 0.531 & \phn19.4  & 0.36 & 197.0 & 31.7 & \phn324  & \phn70              & 0.46 & 1.8      & 140         & 0.72      & 2 \\
    B0834+06 & 1.274 & \phn12.8  & 0.62 & 219.7 & 26.3 & \phn324  & 220                 & 0.69 & 1.2      & 210 &       0.64        & 2 \\
    B1933+16 & 0.359 & 158.5     & 3.70 &  52.4 & -2.1 & 1668     & \phn42              & 0.85 & \phn0.84 & \phn50      & 0.73      & 3 \\
    B0833-45 & 0.089 & \phn69.0  & 0.29 & 263.6 & -2.8 & 1668     & \phn\phn\phn\phn6.2 & 7.6  & 6.4      & \phn\phn7.3 & 0.79-0.87 & 4 \\
\enddata
\tablecomments{
  Columns are as follows:
   (1) pulsar name,
   (2) pulsar period, 
   (3) dispersion measure,
   (4) distance, 
   (5) galactic longitude,
   (6) galactic latitude,
   (7) observing frequency,
   (8) scintillation time,
   (9) scattering time,
   (10) scattering angle,
   (11) decorrelation bandwidth,
   (12) the ratio of distance of the scattering screen to distance of the pulsar,
   (13) the reference where the parameters in columns (10) to (12) were determined.
} 
\tablerefs{(1)\citealt{popov2017}; (2) \citealt{fadeev2018}; (3) \citealt{popov2016}; (4) \citealt{popov2019}. }
\end{deluxetable*}

Of particular interest in our work reported here are investigations of the
influence of scattering on the interferometric visibility function of a
two-element interferometer. The detailed analysis of the substructure of the
visibility function may give us additional information on the characteristics of
the scattering screens.  Early theoretical studies of visibility functions of
two-element interferometers were presented by \citet{goodman1989, narayan1989}.
They distinguished between fast diffractive and slow refractive scintillations
with corresponding time scales of $\tdif$ and $\tref$,
with the diffractive time scale in particular
  related to the size of the diffraction spot,
$\rhodif$.
The physical interpretation of a
visibility function depends on the integration time, $\tint$, and its relation to
the two time scales.  We can distinguish between the snapshot mode when $ \tint
<\tdif$, averaged mode when $\tdif<\tint<\tref$, and ensemble averaging mode
when $\tint>\tref$.  For the meter and decimeter 
 wavelength range typical time
scales, $\tdif$ and $\tref$, for sources in our Galaxy are several minutes for
diffraction scintillations and several weeks for refraction scintillations,
respectively. For our VLBI observations with typical scan lengths of 1000 s, the
visibility function can be measured either in the snapshot mode or the averaged
mode.  Here we present an analysis of the structure of the delay visibility
function for our pulsars 
in the snapshot mode  and average mode and search for characteristics
 that can be related to scattering properties. 

\begin{figure*}[htb]
\includegraphics[width=\textwidth]{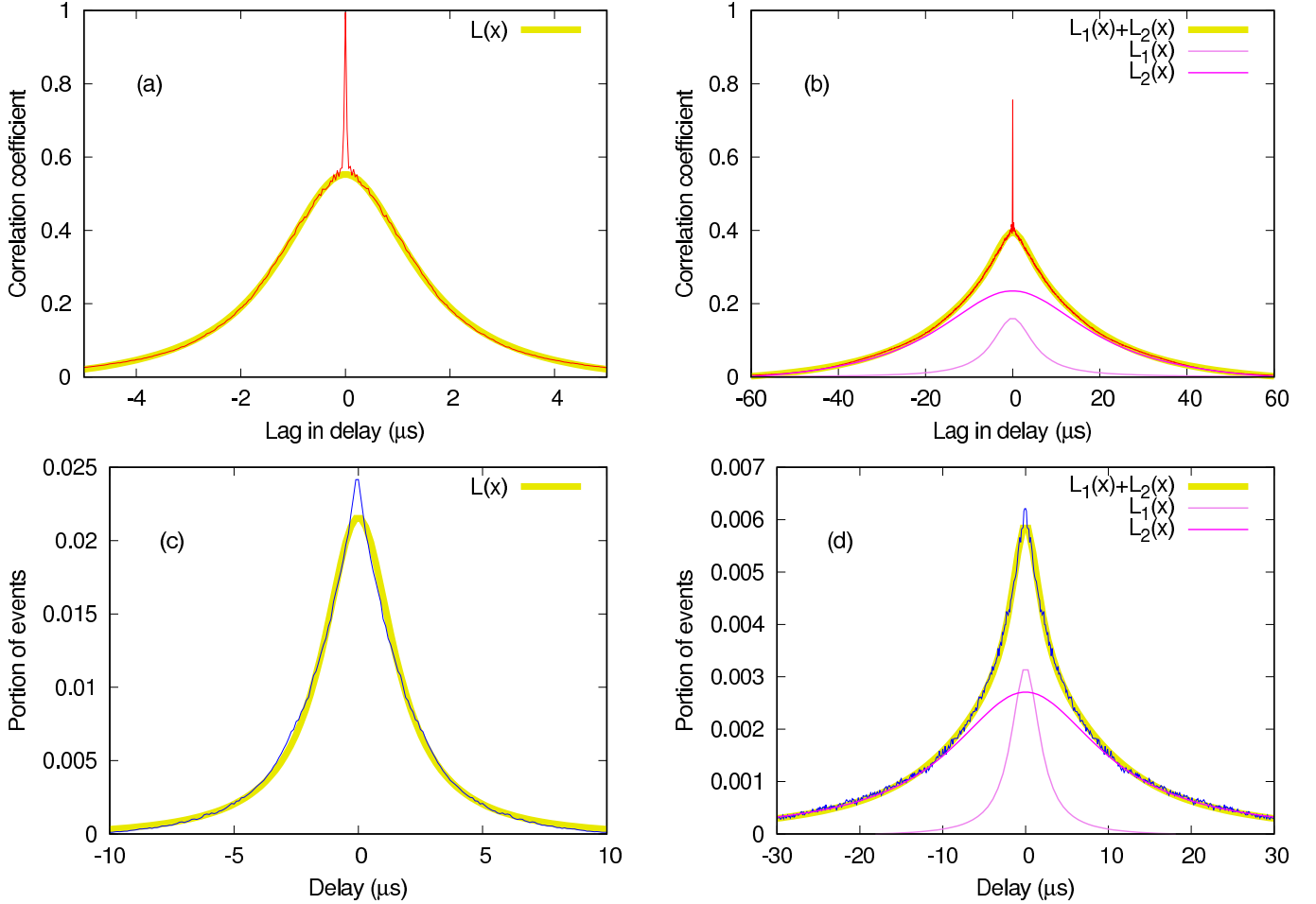}
\caption{ Upper panels: Examples of the function, $CCF^{LR}_{A-B}(\Delta\tau)$,
  for (a) PSR B0823+26 for the baseline GB-WB and for (b) PSR B0833-45 (Vela)
  for the baseline AT-HO given in red. The best fit Lorentzian functions outside
  zero lag in delay are given in yellow.
For PSR B0833-45 (Vela) the best fit is a sum of two Lorentzian functions shown
as violet and magenta lines.  
\\ Lower panels: Results of numerical simulations
of 
distributions of scattered rays in delay
 (c) - assuming a circular scattering
disk, (d) - assuming an elliptical scattering disc with 1:3 axes ratio (see
Sect. \ref{sec:simul} for explanation).
}
\label{fig:CCFfit+simul}
\end{figure*}

\section{Observations and data reduction}  \label {sec:obs} 
Our observations were carried out as part of the scientific program of the
RadioAstron space-VLBI mission \citep{2013ARep...57..153K}.  
For this study we selected pulsar data from
several projects: RAES07a, RAES07b, RAES10a-d, related to the RadioAstron Early
Science Program, RAGS04aj, RAGS04ak, RAGS04al, related to general observing time
proposals, and RAKS02aa, RAKS02as, related to the Key Science Program. The
observation and data reduction parameters are given in Table~\ref{tab:param}.

All our data were processed with the ASC correlator in Moscow with gating and
dedispersion activated \citep{likhachev2017}. The ON-pulse window was centered
on the main component of the average profile, and the OFF-pulse window was
offset from the main pulse by half a period and had the same width as the
ON-pulse window.  The OFF-pulse window was used to establish bandpass correction
for auto and cross-spectra.  The correlator output was sampled synchronously
with the pulsar period. The results of the correlation were given as complex
cross-correlation spectra (cross-spectra) written in standard FITS format.  In
general, the cross-spectra were obtained for each period of the pulsar. Only for
pulsar B0833-45 (Vela) with a very short period (0.0892~s), cross-spectra were
integrated in the correlator over 10 periods, still providing good time
resolution for further analysis.

\begin{deluxetable*}{llrrrllrlhhrl}
\tabletypesize{\footnotesize}
\tablecaption{Parameters of data reduction \label{tab:param}}
\tablewidth{0pt}
\tablehead{
    \colhead{PSR}         & \colhead{Obs. code} & \colhead{Date} & \colhead{$\nu$}  
   & \colhead{$\Tscan$}  & \colhead{$T_\mathrm{tot}$} 
   & \colhead{Pol} & \colhead{$N_\mathrm{ch}$} & \colhead{$\deltatcor$}
   & \nocolhead{$N_\mathrm{vis}$}  & \nocolhead{$M_\mathrm{vis}$} 
   & \colhead{$\Tvis$} & \colhead{Radio telescopes}
 \\
 & & \colhead{(dd.mm.yy)} &\colhead{(MHz)} &\colhead{(s)} 
 & \colhead{(h)} & & & \colhead{(s)} & &  &\colhead{(s)}& \\
    \colhead{(1)} & \colhead{(2)} & \colhead{(3)} & \colhead{(4)} & \colhead{(5)} & \colhead{(6)} &
    \colhead{(7)} & \colhead{(8)} & \colhead{(9)} & \nocolhead{(10)} &
    \nocolhead{(11)} & \colhead{(10)} & \colhead{(11)}
}
  \startdata
    B0329+54 & RAES10,a-d  & 26-29.11.12 & 324 & 570 & \phn4 & LR & 4096 & 0.714 & 48 & 16 & 34.3 & GB\\
    B0823+26 & RAGS04ak,aj & 11.03.14    & 324 & 1170 &17 & LR & 1024 & 0.5306 & 64 & 34 & 34.0 & GB,WB\\
    B0834+06 & RAGS04al    & 08.04.15    & 324 & 1170 &\phn1.5 &  LR & 65536 & 1.2737 & 114 & 8 & 145.2 & AR,GB,WB\\
    B1933+16 & RAKS02aa    & 01.08.13    & 1668 & 570 &\phn1.5 &  R  & 2048 & 0.3587 & 88 & 18 & 31.6 & AR,TR,SV\\
    B0833-45 & RAKS02as    & 15.12.13    & 1668 & 1170 &\phn2.5 & LR & 8192 & 0.9830 & 8 & 148 & 7.9 & AT,HO,CD,HH\\
    B0833-45 & RAES07a     & 10.05.12    & 1668 & 570 &\phn3.0 & LR & 8192 & 0.9830 & 8 & 70 & 7.9 & PA,MP,TI,HH,HO\\
    B0833-45 & RAES07b     & 18.05.12    & 1668 & 570 &\phn1.5 & LR & 8192 & 0.9830 & 8 & 70 & 7.9 & PA,AT,HO,MP,HH\\
    \enddata
   \tablecomments{
  Columns are as follows:
   (1) pulsar name,
   (2) code of the experiment, 
   (3) date of observations,
   (4) observing center frequency,
   (5) duration of observing scan in seconds,
   (6) total observing time in hours,
   (7) circular polarization, left hand, LCP, L, right hand, RCP, R,
   (8) number of channels used in the correlator,
   (9) sampling time of the correlator output in seconds,
   (10) $T_\mathrm{vis}$ - time in seconds for visibility calculation,
   (11) earth radio telescopes: AR -Arecibo, GB - Green Bank, 
        AT - Australia Telescope Compact Array, HO - Hobart, 
        HH -  Hartebeesthoek, CD - Ceduna, MP - Mopra, SV - Svetloe, 
        TI - Tidbinbilla, PA - Parkes, WB - Westerbork. 
}
\end{deluxetable*}

At the next stage, we retrieved the results of correlation processing from the
FITS files using the CFITSIO package \citep{1999ASPC..172..487P} and computed the
fringe visibility magnitude $|V_{A-B}(\tau,f)|$ as a function of
delay, $\tau$,
and fringe rate, $f$, for every time interval of duration $\Tvis$, with $\Tvis<\tscint$ by  using 
$\Tvis/\deltatcor$ consequent complex cross-spectra from the correlator output for each
two-element interferometer with stations A and B.
Then for each $\tau$ we determined the 
fringe rate that maximizes $|V_{A-B}(\tau,f)|$.  Not surprisingly, these
  fringe rates were always close to zero.  For further analysis we used the
  cross section, $|V_{A-B}(\tau,f_{\max})|=|V_{A-B}(\tau)|$ at $f_{\max}$.  For
  every scan of duration $\Tscan$, we obtained $\Tscan/\Tvis$ such cross
  sections of the visibility function, $|V_{A-B}(\tau)|$.  We call this set of
  cross sections of the visibility magnitudes, which are consecutive in time, t,
  the dynamic visibility magnitude, $|DV_{A-B}(\tau,t)|$.

Our goal was to probe these functions in detail, study their
  characteristics as a function of projected baseline length where possible and
  extract scintillation parameters from them. Since these functions were
  relatively noisy and were not appropriate for obtaining single scintillation
  parameter values, we used two-dimensional cross correlation functions (CCFs)
  and autocorrelation functions (ACFs) to improve the signal to noise ratio.

We distinguish between one-baseline correlations and two-baseline
  correlations of interferometer observations. In particular, for one-baseline
  correlations of interferometer observations, we compute the two-dimensional
  cross correlation functions $2dCCF^{LR}(\Delta\tau,\Delta t)$ between the LCP
  and RCP polarization channels of $|DV_{A-B}(\tau,t)|$. First, we subtracted
  the mean level in every $|DV_{A-B}(\tau,t)|$ determined "off-spot",
  i.e. outside the region of increased values of $|V_{A-B}(\tau)|$.  Then we
  computed the cross-correlation functions. \footnote{For technical reasons,
    instead of cross correlating the two functions directly, we derived the
    cross correlation by using the cross-correlation theorem. This procedure
    simplified the computation in our case.  We computed the functions $2dCCF$
    as the inverse Fourier transform of the product of the two-dimensional
    complex cross-spectrum of $|DV^R_{A-B}(\tau,t)|$ in RCP and the
    two-dimensional complex cross-spectrum of $|DV^L_{A-B}(\tau,t)|$ in LCP. In
    order to avoid cyclic convolution inherent in the Fourier transform, we
    expanded the functions $|DV^R_{A-B}(\tau,t)|$ and $|DV^L_{A-B}(\tau,t)|$ by
    zero values twice in both coordinates.} The resulting functions,
  $2dCCF^{LR}(\Delta\tau,\Delta t)$, were then normalized by the corresponding
  $2dACF$, that is by $\sqrt{2dACF^{L}(\Delta\tau=0,\Delta t=0) \times
    2dACF^{R}(\Delta\tau=0,\Delta t=0)}$.

For the observations of B1933+16 and for part of them of PSR B0833-45
  (Vela), we recorded only LCP or RCP and therefore considered for our further
  analysis of one-baseline correlations only the autocorrelation functions,
  $2dACF^{L}(\Delta\tau,\Delta t)$ or $2dACF^{R}(\Delta\tau,\Delta t)$ instead
  of the cross correlation function between the RCP and LCP channels.

The last functions to mention concern two-baseline correlations of
  interferometer observations. Here we measure similarities between the output
  of, for instance, a short baseline interferometer and output of a long
  baseline interferometer to obtain information about the difference in the
  diffraction pattern the two interferometers observe. The resulting functions
  are $2dCCF^{LL}_{(A-B)\times(C-D)}$ and $2dCCF^{RR}_{(A-B)\times(C-D)}$ which
  are the cross-correlations between $|DV^L_{A-B}(\tau,t)|$ and
  $|DV^L_{C-D}(\tau,t)|$ for the two baselines, AB and CD, for the LCP channel
  and the equivalent correlations for the RCP channel.  At the heart of our
  analysis are the cross sections of these functions at $\Delta t=0$. These are
  the one-dimensional functions, $CCF^{LR}_{A-B}(\Delta\tau)$,
  $ACF^{L}_{A-B}(\Delta\tau)$, and $ACF^{R}_{A-B}(\Delta\tau)$ and
  $CCF^{L}_{(A-B)\times(C-D)}(\Delta\tau)$ and
  $CCF^{R}_{(A-B)\times(C-D)}(\Delta\tau)$.
\section{Results}   \label{Results}
Our results are obtained from the five one-dimensional CCF and ACF
  functions described at the end of the previous section.  Typical examples of,
  for instance, the function, $CCF^{LR}_{A-B}(\Delta\tau)$, for the baseline
  GB-WB for PSR B0823+26 and for the baseline AT-HO for PSR B0833-45 (Vela) are
  given in Figure \ref{fig:CCFfit+simul}(a, b).
The function consists of an
unresolved spike and a broad component. We describe each in turn, 
give examples
of these functions for all five pulsars. and then describe parameter estimates.
\subsection{The unresolved spike}
Almost  all of our three CCF and all of our two ACF functions,  consist of an unresolved spike at zero
delay lag, that is at $\Delta \tau = 0$, and a smoothly and slowly varying envelope
(SVE) starting at an amplitude of approximately half the amplitude of the spike
and extending to several $\mu$s in negative and positive delay lags.
\begin{figure}[htb]
\includegraphics[width=80mm]{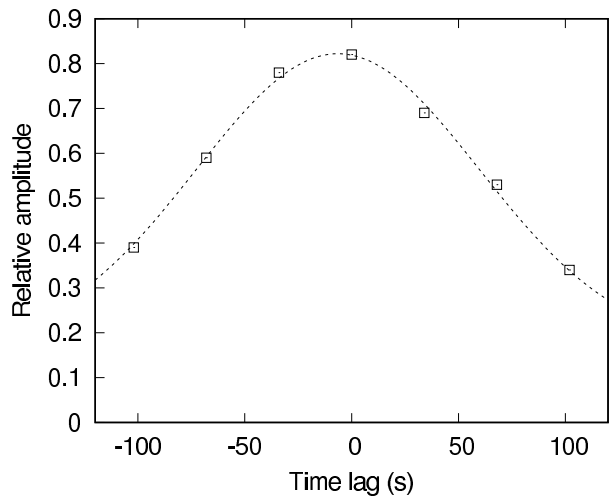}
\caption{Relative amplitude of the unresolved spike versus time lag, $\Delta t$,
  in our two dimensional ACFs for PSR B0329+54. The dashed line corresponds to a
  fit with a Gaussian.}
\label{fig:spike_vs_time}
\end{figure}
\begin{figure*}[htb]
\includegraphics[width=\textwidth]{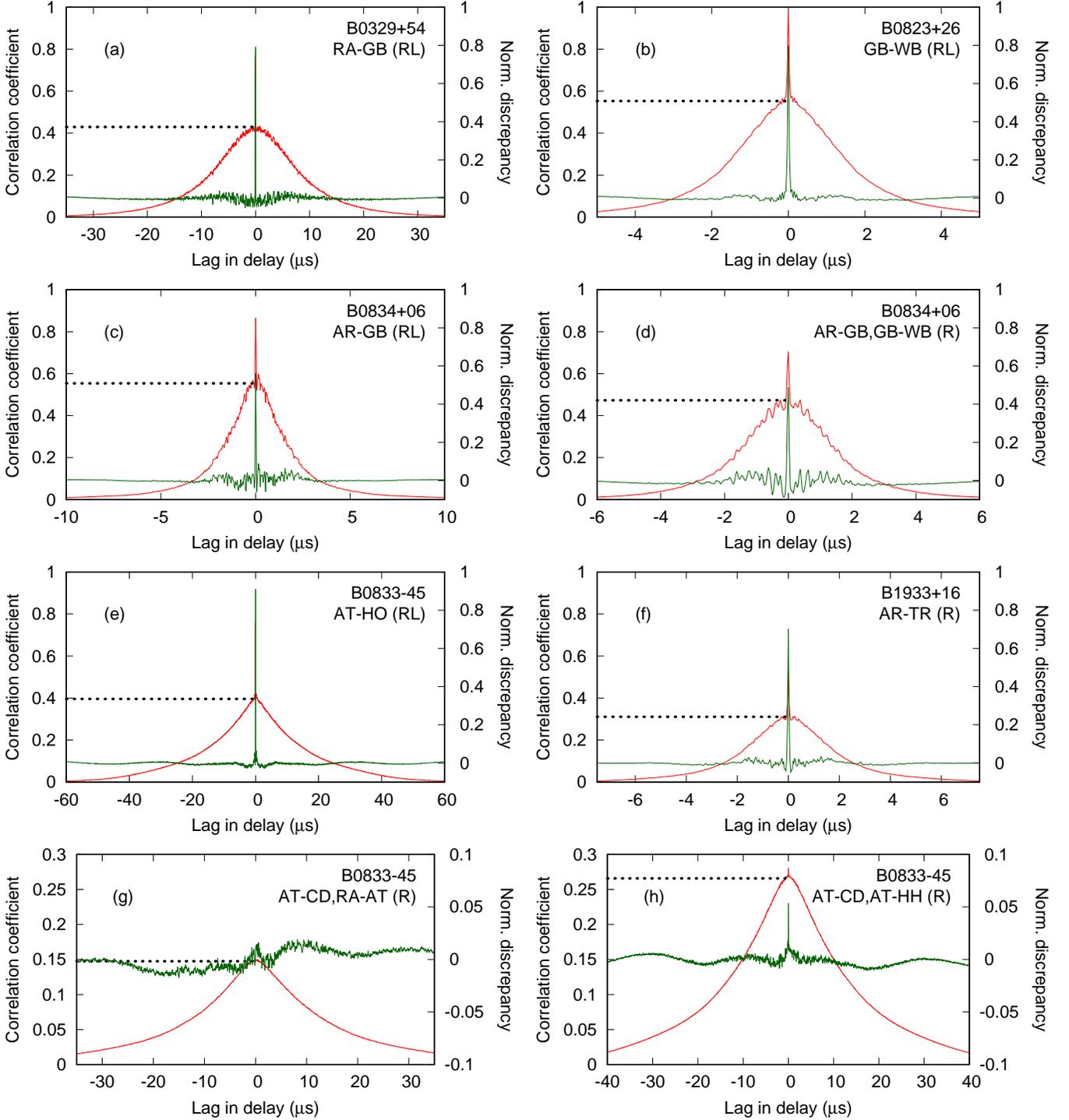}
\caption{Examples of one-dimensional CCF's and ACF's of visibility amplitudes.
  The pulsar names, the baselines and the polarization information are given in
  each panel.  
 Observed values are plotted by red lines.
    Dotted lines indicate the values of $L(0)$, where
    $L(\Delta\tau)$ is the best Lorentzian fit of a broad smooth part of
    measurements obtained outside of the narrow spike at $\Delta\tau=0$.  
    Green lines show the discrepancy between measured values and
    $L(\Delta\tau)$.   
    The discrepancy was normalized by $L(0)$ in order to facilitate the comparison
    with predictions of the AMN model.  For all pulsars but PSR B0833-45 (Vela)
    the observations are well approximated by a Lorentzian function. For Vela
    the sum of two Lorentzians were needed for the fit.  Note, that in two
    bottom panels the scale of the vertical axes is enlarged.
}
\label{fig:delayCCF}
\end{figure*}
These characteristics are reminiscent of the ACF for pulsar microstructure
consisting of an unresolved spike at zero time lag and a broader component due
to the fast intensity fluctuations of the radio emission of pulsars on the time
scale of 10's to 100's of microseconds.  These characteristics were interpreted
by \citet{rickett1975} in terms of the amplitude-modulated noise model (AMN).

Since the individual visibility functions were computed in the snapshot mode, we
interpret the fine structure of $|V_{AB}(\tau)|$ in this mode to be band-limited
white noise. We call it scintillation noise (SN). The amplitude of the SN is
decreasing with increasing magnitude of the time lag, $\Delta t$ in our two
dimensional ACFs, $2dACF^L(\Delta\tau, \Delta t)$ and $2dACF^R(\Delta\tau, \Delta t)$, 
as demonstrated in Figure \ref{fig:spike_vs_time} for the
pulsar B0329+54.  Such behavior can be fit by a Gaussian. Thus, we estimated the
scintillation time, $\tscint=115 s$, as the $1/e$ half width of this curve. The
scintillation time is in approximate agreement with the value of
$\tscint=110-112 s$ determined earlier for this pulsar from single-dish
autocorrelation spectra \citep{popov2017}. We find the same characteristic of
the SN amplitude decreasing with increasing magnitude of time lag, $\Delta t$,
approximately as a Gaussian for each pulsar in our sample and list our values
for $\tscint$ in Table~\ref{tab:param0}.
\subsection{The broad component}        \label{sec:broad}
Apart from the unresolved spike, all our one-dimensional CCFs and ACFs are
characterized by a broad component with a slowly varying envelope (SVE) starting
at an amplitude of approximately half the amplitude of the unresolved spike at
zero delay lag and extending to several $\mu$s in negative and positive delay
lags.
Following \citet{popov2016}'s example of PSR B1749-28, we fit the shape of the
SVE by a Lorentzian function 
\begin{equation}                                          \label{eq:Ldef}
L(\Delta\tau)= rw/(\Delta\tau^2+w^2)+C\,.  
\end{equation}
We allow
for a constant, $C$, to compensate for a possible offset in $|DV_{A-B}(\tau, t)|$.
\footnote{
In principle the constant $C$ should be equal to zero if the
      baseline of the individual visibility magnitudes, $|V_{A-B}(\tau)|$ could
      be exactly known and accurately subtracted. However, errors in the
      determinations of the baselines migrated into the construction of the
      dynamic visibility magnitude, $|DV_{A-B}(\tau,t)|$, and therefore into the
      CCFs and the ACFs under consideration here. The constant, $C$, eliminated
      this influence in the fit.
}
The maximum of the function above the constant, $C$, is
$A=r/w$ at $\Delta\tau=0$, and the half-width at half maximum ($HWHM$) of
the function above $C$ is $w$.

Temporal smearing of a pulse due to scattering in the ISM is
  conventionally characterized by the scattering time, $\tausc$.  
  As we show in Appendix A, $\tausc \approx w/2$ for the Lorentzians of
  equation (\ref{eq:Ldef}) fit to our CCFs and ACFs. 
  While the determination of $\tausc$ through
  temporal smearing of pulses is often difficult and can be prone to relatively
  large uncertainties, our new method of determining $\tausc$ is relatively easy and
  can provide more accurate values.
We determine the parameters through least-squares
fitting. In Figure~\ref{fig:CCFfit+simul}(a) we show the fit with a yellow line.
For B0833-45 (Vela) it was obvious that a Lorentzian  would not provide
for a good fit.  Instead, the sum of two Lorentzians with different parameters were
needed.  We discuss the special case of PSR B0833-45 (Vela) in more detail
below.  We show the fits in Figure~\ref{fig:CCFfit+simul}(b).
\subsection{Typical examples of CCFs and ACFs for all five pulsars}
In Figure~\ref{fig:delayCCF} we present examples of the CCFs and ACFs and the
results of the Lorentzian fits for all the pulsars in our sample.  However,
instead of plotting the fit Lorentzians as in Figure~\ref{fig:CCFfit+simul}, we
plot the difference between the measured
CCFs or ACFs and the fit Lorentzians to indicate
visually more clearly the goodness of the fit. 
Furthermore, we normalize the difference by the
value L(0), which is shown in Figure~\ref{fig:delayCCF} by horizontal dotted lines.
The normalization permits to compare directly the observed values of CCFs and ACFs
at $\Delta\tau=0$ with the value of $2L(0)$ predicted by the AMN model.
  In the 
  panels, (a-c), (e) and (f) we show the CCFs and in one case the ACF for
  one baseline only. Assuming only a minor influence of cross correlating the RCP
  with the LCP channel instead of autocorrelating one polarisation only, these
  five plots should have similar characteristics concerning the unresolved
  spike.  In panels (d), (g), and (h) we show the CCF's for two baselines. In all our
  plots, for B0329+54, B0823+26, B0834+06, and B1933+16, one Lorentzian alone
  fits the CCFs and the ACF well. For PSR B0833-45 (Vela), however, as already
  indicated in Figure~\ref{fig:CCFfit+simul}, clearly two Lorentzians were
  needed.

Focussing first on the maximum of the unresolved spike, it is clear from
  the results for single-baseline CCFs and ACF  that the amplitude is
  always higher than 0.6 of the maximum of the Lorentzian, reaching  0.95, which is
  almost the predicted value of 1.0 for the AMN model. The situation is different for
  the CCF's between data from different baselines. In all three cases the
  amplitude is lower than 0.6. In particular for PSR B0833-45 (Vela) the
  amplitude decreases to 0.05 for the two earth-earth baselines and goes to zero
  for earth-earth to earth-space baselines.

In contrast to the unresolved spike, for the broad component of all but PSR
0833-45 (Vela), the residuals show only very small deviations from statistical
noise, indicating the excellent quality of the fit with one Lorentzian. For the
broad component of B0833-45 (Vela), where two Lorentzians were needed, the
residuals are somewhat larger but still indicate a good fit.

\subsection{Parameter estimates}
For all the pulsars apart from PSR B0833-45 (Vela), the results are given in
Tables \ref{tab:res1}.  For each of the four pulsars we list the correlated
baselines as a function of increasing projected baseline length together with
 the corresponding interferometric angular resolution,
$\thetares$, in units of the angular scattering angle, $\thetasc$,
 and $w$, the HWHM of the Lorentzian fits.  
A scattering disk is resolved  when $\thetares/\thetasc<1$.

\begin{deluxetable}{cclccl}
\tabletypesize{\footnotesize}
\tablecaption{Results of the Lorentzian fits for four pulsars \label{tab:res1}}
\tablewidth{0pt}
\tablehead{ \colhead{PSR} & \colhead{Corr. baselines} 
   & \colhead{Function} & \colhead{Length}   
   & \colhead{$\thetares/\thetasc$}& \colhead{$w$}\\
                          &                          &                & \colhead{($M\lambda$)} & &  \colhead{($\mu$s)}   \\
            \colhead{(1)} & \colhead{(2)}            & \colhead{(3)}  & \colhead{(4)}     & \colhead{(5)} & \colhead{(6)}
         }
\startdata
 B0329+54 & RA-GB & $CCF^{LR}$         & \phn65 & \phn0.66 & 8.0(1) \\
 &                  RA-GB  &$CCF^{LR}$ & \phn98 & \phn0.44 & 8.3(2) \\
 &                  RA-GB  & $CCF^{LR}$ & 190   & \phn0.23 & 8.4(1) \\
 &                  RA-GB & $CCF^{LR}$ & 235    & \phn0.18 & 8.5(1) \\
 \hline
 \\
 B0823+26 & GB-WB &$CCF^{LR}$  & \phn\phn\phn6.5 & 17\phn\phn\phn  & 1.48(8) \\
 &                  RA-GB  & $CCF^{LR}$ & \phn51 & \phn2.2\phn     & 0.56(6) \\
 &                  RA-GB  & $CCF^{LR}$ & \phn55 & \phn2.0\phn     & 0.74(7) \\
 &                  RA-GB  & $CCF^{LR}$ & \phn61 & \phn1.8\phn     & 0.87(9) \\
 \hline
 \\
 B0834+06 & AR-GB &$CCF^{LR}$          & \phn\phn\phn2.5 &  66\phn\phn\phn & 1.6(1) \\
 &                  AR-WB &$CCF^{LR}$  & \phn\phn\phn6.0 &  28\phn\phn\phn & 1.5(1) \\
 &                  GB-WB & $CCF^{LR}$ & \phn\phn\phn6.0 &  28\phn\phn\phn & 1.5(1) \\
 &                  RA-AR & $CCF^{LR}$ & 165             & \phn2.0\phn     & 1.3(1) \\
 &                  RA-GB & $CCF^{LR}$ & 165             & \phn2.0\phn     & 1.3(2) \\
 \hline
 \\
 B1933+16 & AR-TR & $ACF^{R}$  &  \phn36                        & \phn5.7\phn & 1.7(1) \\
                  & AR-SV  & $ACF^{R}$  &   \phn36              & \phn5.7\phn & 1.8(1) \\
                  & (AR-TR)$\times$(AR-SV) &$CCF^{RR}$ & \phn36 & \phn5.7\phn & 1.8(2) \\
                  & RA-AR & $ACF^{R}$               & 34-164    & 6.0-1.2     & 1.5(2) \\
\enddata                        
\tablecomments{
  Columns are as follows:
   (1) pulsar name,
   (2) two-element interferometer with stations as in Table 2,
   (3) the function analyzed, 
   with $CCF^{LR}=CCF^{LR}_{A-B}(\Delta\tau)$,
     $ACF^{R}=ACF^{R}_{A-B}(\Delta\tau)$, and
     $CCF^{RR}=CCF^{R}_{(A-B)\times(C-D)}(\Delta\tau)$, with stations, A, B, C,
     D, as defined in section 2.
   (4) length of projected baseline in millions of wavelengths, $M\lambda$,
   (5) interferometer angular resolution given by projected baseline
     length from (4) in units of the angular scattering angle, $\thetasc$,
   (6) HWHM of a Lorentzian fit to the function analyzed,
   the number in
     parentheses is the approximate error (1$\sigma$) in the last digit of $w$
     computed from the rms variation and the number of scans during the
     observation time, $T_\mathrm{tot}$, assuming Gaussian statistics.
}
\end{deluxetable}

 The formal uncertainty of our estimated
 values for $w$ for a single scan is about 1-3\%, 
 while the peak to peak variation between successive scans is about 10\%, 
 which reflects the variation due to scintillation.
The values of $w$ are averages over the whole observing time of about 
one to
a few hours, and have to be considered as obtained in the average mode of
observation.  There is a hint that for all pulsars but PSR B0329+54, $w$ is
decreasing with increasing baseline projection, while the scattering disk
becomes more and more resolved by the beam of the two-element interferometer.
Such behavior was predicted theoretically by \citet{gwinn1998}.

 The anomalous dependence of $w$ on baseline projection,
  $|b|$, for PSR B0329+54 can perhaps be explained by rapid changes in the
  properties of the scattering screen that mask the effect of variable
  baseline.  Observations of the pulsar by
  \citet[Fig. 4]{1999ApJS..121..483B} show that at 327 MHz the
  decorrelation bandwidth may change by a factor of two over one or
  two days.

 It is also possible that the weak dependence of $w$ on
  baseline projection for PSR B0329+54 is a consequence of the fact that the ratio
  $\thetares/\thetasc$ is smaller than unity and also much lower than for other
  pulsars. It appears that $w$ decreases with increasing baseline projection as
  long as 
  scattering disk is resolved, $w$ looses its dependence on baseline projection
  and becomes constant. In this context it is interesting to note that such
  behavior was indeed found for the second moment of visibility by Gwinn et
  al. (1998). It remains to be seen whether a similar behavior can also be
  derived for $w$.

The visibility characteristics of PSR B0833-45 (Vela) are more
complex and we, therefore, give our results separately in Table
\ref{tab:resVela}.  The pulsar was observed three times between
2012 and 2013, and as in Table \ref{tab:res1}, we list the
sessions and dates together with the baselines, the polarization
and the projected baseline lengths in the order of increasing length,
as well as the angular resolution in units of the scattering angle.
For observing dates 10 May 2012 and 15 December 2013 we needed two
Lorentzians to fit the cross-sections of the CCFs and ACFs, with
HWHM, $w_1$ and $w_2$, of about $4-8~\mu s$ for the short time
scale and $15-25~\mu s$ for the long time scale, respectively.  
In contrast, for the observing date of 18 May 2012, which is about one orbit
after the date of 10 May 2012, one Lorentzian only was sufficient for the
fit. 
This change is particular striking for the baseline MP-HO with the same
projected baseline length and the same position angle, PA, but for 8 days apart.
We list the values for $w_1$ and $w_2$, together with the values for
the amplitudes of the Lorentzians as well as the position angles
of the baselines also in Table \ref{tab:resVela}.  In contrast to
B0823+26, B0834+06, and B1933+16 there is no decrease of 
either $w_1$ or $w_2$ with increasing projected baseline
length.  Further, there is also no apparent dependence of the
amplitude on the baseline position angle.

We now focus on the correlation of the scintillation noise (SN)
between different baseline projections in more detail. In particular, we compare
the height of the unresolved spike at zero delay lag in
Figures~\ref{fig:delayCCF}(d,g,h). The correlation of SN between
short baselines as in Figure~\ref{fig:delayCCF}(f) with projected
baseline lengths of 2.5 and 6.0 M$\lambda$ is relatively high. The
correlation decreases significantly between short and
intermediately long baselines with lengths of 5.8-7.5 and 54
M$\lambda$ as in Figure~ \ref{fig:delayCCF}(h) and completely
vanishes between short and long (earth-space) baselines of
5.8-7.5 and 630~M$\lambda$ as in Figure \ref{fig:delayCCF}(g).
These characteristics appear to be related to the size of a
diffraction spot, $\rhodif$, in the scattering screen relative to
the difference of the projected lengths of the pair of the
correlated baselines, $B_{A-B}$ and $B_{C-D}$.  If
$B_{A-B}<\rhodif$ and $B_{C-D}<\rhodif$, then the two
interferometers observe the same diffraction spot with about the
same angular resolution and consequently for SN, the correlation
is relatively high. If $B_{A-B}<\rhodif$ and $B_{C-D}\sim\rhodif$,
the correlation decreases. In the extreme case, $B_{A-B}<\rhodif$
and $B_{C-D}>\rhodif$, the correlation completely vanishes since
the two interferometers observe different diffraction spots.

\begin{deluxetable*}{lllrcclllr}
\tabletypesize{\footnotesize}
\tablecaption{Results of the Lorentzian fits for PSR B0833-45 (Vela) \label{tab:resVela}}
\tablehead{ \colhead{Session} & \colhead{Corr. baselines}         & \colhead{Function} & \colhead{Length} 
& 
    \colhead{$\thetares/\thetasc$}
   &\colhead{$w_1$}
       & \colhead{$w_2$} &   \colhead{$A_1$} & \colhead{$A_2$} &  \colhead{PA} \\
  & & &\colhead{($M\lambda$)}  &\colhead{} & \colhead{($\mu s$)} & \colhead{($\mu s$)}  &\colhead{} 
  &\colhead{} &\colhead{($^{\circ}$)} \\
 \colhead{(1)} & \colhead{(2)} & \colhead{(3)} & \colhead{(4)} & \colhead{(5)} & 
 \colhead{(6)} & \colhead{(7)} & \colhead{(8)} & \colhead{(9)} & \colhead{(10)}
}
\startdata
RAES07a     &  MP-PA & $ACF^{L}$   & 1.0      & \phn30\phn\phn\phn & \phn4.2(2) & 15.5(4) & 0.05 & 0.14 & 50\\
10 May 2012 &  TB-PA & $ACF^{L}$   & 1.5      & \phn20\phn\phn\phn & \phn5.7(6) & 21.9(8) & 0.12 & 0.31 & 11\\
            &   TB-HO & $ACF^{L}$  & 4.4      & \phn7.0\phn        & \phn4.1(3) & 16.5(8) & 0.05 & 0.16 & 36\\
            &   PA-HO &$ACF^{L}$   & 5.8      & \phn5.2\phn        & \phn5.0(2) & 16.6(7) & 0.0011 & 0.0012 &  30\\
            &   MP-HO &$ACF^{L}$   & 6.7      & \phn4.6\phn        & \phn5.6(2) & 18.0(2) & 0.006 & 0.007 &  33 \\
            &   TB-RA & $ACF^{L}$  & 725\phn  & \phn0.04           & 12.0(4)    & & 0.035 & & 166\\[2ex]
           &    (TB-PA)$\times$(MP-HO) &$CCF^{LL}$ &1.5, 6.7 &     & \phn5.9(2) & 17.2(5) & 0.024 & 0.035&  \\
           &    (TB-PA)$\times$(MP-PA) &$CCF^{LL}$ &1.5, 1.0 &     & \phn5.9(1) & 19.2(5) & 0.089 & 0.187&  \\
           &    (TB-PA)$\times$(MP-TB) &$CCF^{LL}$ &1.5, 2.4 &     & \phn6.3(2) & 23.2(6) & 0.14 & 0.32 &  \\
           &    (TB-RA)$\times$(MP-TB) &$CCF^{LL}$ &725, 2.4 &     & \phn6.4(2) & 17.4(4) & 0.032 &0.090&  \\
\hline
\\
RAES07b      &    AT-MP &$CCF^{LR}$ & 0.5 & 61\phn\phn      & 17.1(4) & & 0.32  & & 45\\ 
18 May 2012  &   PA-MP& $CCF^{LR}$  & 1.0 & 30\phn\phn      & 18.0(4) & & 0.39  & & 50\\
             &  PA-AT & $CCF^{LR}$  & 1.5 & 20\phn\phn      & 20.5(5) & & 0.44  & & 45\\
             & PA-HO &$CCF^{LR}$    & 5.8 & \phn5.2         & 13.8(4) & & 0.14  & & 30\\ 
             &  MP-HO & $CCF^{LR}$  & 6.7 & \phn4.6         & 11.4(4) & & 0.004 & & 33\\
             &  AT-HO &$CCF^{LR}$   & 7.3 & \phn4.3         & 12.4(3) & & 0.08  & & 33\\
             &  AT-HH &$CCF^{LR}$   &52.2 & \phn\phn0.58    & 19.6(5) & & 0.04  & & 90\\
             &  AT-RA & $CCF^{LR}$  &1065\phn &\phn\phn0.03 & 14.0(3) & & 0.017 & & 166\\[2ex]
             &  (PA-AT)$\times$(AT-MP) & $CCF^{RR}$ &  1.5, 0.5  & & 18.8(4) & & 0.39 & &  \\
             &  (AT-HH)$\times$(PA-HH) & $CCF^{RR}$ & 52.2, 52.2 & & 20.1(5) & & 0.43 & & \\
\hline
\\
RAKS02as          &  AT-CD &$CCF^{LR}$  &  5.8-7.5 & 5.2-4.1  & \phn5.0(2) & 15.9(3) & 0.22 & 0.30 &  112-150\\
15 December 2013  &  AT-HO & $CCF^{LR}$ &  6.6-7.3 & 4.6-4.3  & \phn4.4(1) & 19.7(3) & 0.14 & 0.28 &  33-46\\
                  &  AT-HH & $CCF^{LR}$ & 54\phn  & \phn0.55  & \phn7.8(1) & 22.0(4) & 0.15 & 0.26 &  80-92\\
                  &  AT-RA &$CCF^{LR}$  & 630\phn  & \phn0.05 & \phn8.6(2) & 25.0(4) & 0.08 & 0.06 &  130\\[2ex]
  &  (AT-CD)$\times$(HO-CD) & $CCF^{RR}$   & 5.8-7.5, 9.3 &    & \phn1.1(1) & 10.0(2) & 0.03 & 0.05 & \\
  &  (AT-CD)$\times$(AT-HO) & $CCF^{RR}$   & 5.8-7.5, 6.6-7.3& & \phn3.3(1) & 17.4(3) & 0.20 & 0.10 &  \\
  &  (AT-CD)$\times$(AT-RA) & $CCF^{RR}$   & 5.8-7.5, 630 &    & \phn6.0(2) & 17.9(3) & 0.06 & 0.09 &  \\
  &  (AT-CD)$\times$(AT-HH) &$CCF^{RR}$    & 5.8-7.5, 54  &    & \phn6.6(2) & 19.3(4) & 0.11 & 0.17 &  \\
  &  (AT-HH)$\times$(AT-RA)&$CCF^{RR}$     & 54, 630      &    & \phn7.3(2) & 18.3(4) & 0.07 & 0.10 & \\
\enddata
\tablecomments{
Columns are as follows:
 (1) - Session code and date,
 (2) - designation of baseline  or baseline combination,
 (3) - the function analyzed, for definition, see Table 3,
 (4) - length of baseline projection in millions of wavelengths, $M\lambda$,
 (5) interferometer angular resolution given by projected baseline
   length from (4) in units of the angular scattering angle, $\thetasc$,
 (6,7) - HWHM of a Lorentzian fit of the function in column (3) where in the
 majority of cases a sum of two Lorentzians with HWHM $w_1$ for the narrow
 Lorentzian, and HWHM $w_2$ for the broad Lorentzian was needed to fit the shape
 of the slowly varying envelope (SVE), 
 errors in parentheses are defined as in Table 3,
 (8,9) amplitudes corresponding to the Lorentzians with $w_1$ and $w_2$, respectively,
 (10) - position angle of baseline projection in degrees.
}
\end{deluxetable*}
\section{Numerical simulation of scattering}    \label{sec:simul}
Our analysis of the 
 CCFs and ACFs 
 has shown that
for the set of our pulsars a single Lorentzian was sufficient for
the fit, except for B0833-45 (Vela) for which 
in many cases
 the sum of two
Lorentzians was needed. Through a numerical simulation, we show
that the difference can be interpreted in terms of circularly
symmetric and non-circularly symmetric scattering of the radio
radiation in the inhomogeneities of the scattering screen.  

We consider the probability distribution in delay for scattered rays refracted
on a thin screen.  For the small angle approximation, the geometric time delay,
$\tau$, is given as a function of the scattering angle, $\theta$, as
$\tau=\theta^2d_\mathrm{eff}/(2c)$, where $d_\mathrm{eff}=Dd/(D-d)$, with $D$
and $d$ as distances to the pulsar and the screen, respectively
\citep{1993ApJ...410..673G}.  Let the screen contain $n$ refractors with
coordinates $x_i$,~$y_i$ ($1\leq i\leq n$), selected from a two-dimensional
Gaussian distribution with standard deviations corresponding to major and minor
axes equal to $\sigma_x^2$ and $\sigma_y^2$, respectively. We compute the mutual
geometrical delays between each of the rays as $\tau_{ij}=\theta_i^2-\theta_j^2$
with $n(n-1)$ combinations. Since $\theta^2\propto \rho^2=x^2+y^2$, we compute
the delays as $\tau_{ij}=\rho_i^2-\rho_j^2$. In order to achieve a smooth
distribution, we assume $n=100$, and we average 100 simulations.

In Figure \ref{fig:CCFfit+simul} (c) we show the distribution of
computed delays for a circular ($\sigma_x=\sigma_y$)
scattering disk, and in Figure \ref{fig:CCFfit+simul} (d) for an
elliptical scattering disk ($\sigma_y=3\sigma_x$). For the circular
disk the distribution is well fit by a single Lorentzian, whereas
for the elliptical disk the sum of two Lorentzians is needed.
 The distributions shown in Figure 1 (c,d) reflect the shape of the average visibility
 function in delay. As it is explained in the Appendix, the ACF of a Lorentzian is
 also a Lorentzian with the HWHM  twice as large as that of the original
 Lorentzian. Similarly, for our cases, the CCFs of the Lorentzians we consider
 are also, at least approximately, Lorentzians, although that is harder to show
 as explained in the Appendix.
 Despite the simplicity of our model, we found a good
 correspondence between the simulated distributions and
 the SVEs of the CCF's and ACF's obtained in our analysis of the substructure
of visibility functions for pulsars. Our analysis
is valid for an interferometer with a short baseline when a
scattering disk is not resolved.  The scale on time delay in the
bottom panels of Figure \ref{fig:CCFfit+simul} is arbitrary, it
depends on the values of the 
 distance to the pulsar $D$, the 
 distance to
the screen $d$, and the 
 scattering angle $\theta$.

\section{Discussion} \label{sec:disc}
The main result of our analysis is that Lorentzians fit the
SVE in our CCF's and ACF's, and therefore in general the envelope of the
two-element interferometer delay visibility functions very well, and that
numerical simulations of scattering confirm that Lorentzians are
indeed expected. Isotropic scattering in the plane of the sky
results in a single Lorentzian for 
the SVE of the CCF's and ACF's
with a HWHM twice as large as $\tausc$, whereas anisotropic
scattering with the scattering disk being elliptical results in
two Lorentzians with different HWHMs. 
In fact, since we found that Lorentzians can be fit for all our five pulsars
with a variety of pulsar parameters, it is likely that this is a general
characteristic of all pulsars that the delay visibility functions can be
described by one or possibly more Lorentzians depending on the complexity of the
scattering medium.
Our pulsars B0329+54,
B0823+26, B0834+06 and B1933+16 are all undergoing isotropic
scattering. Their galactic coordinates and the relative distances
of their scattering screens (see, Table ~\ref{tab:param0})
indicate that for the first three pulsars the screens are
approximately associated with the Local Arm in our Galaxy and for
B1933+16 with the Carina Sagittarius Arm (see also,
\cite{fadeev2018,popov2017}). Despite the large range of their
distances, $D$, dispersion measures, $DM$, scintillation times,
$\tscint$, scattering times, $\tausc$, scattering angles,
$\thetasc$, and decorrelation bandwidths, $\Deltafdif$, no
dependence on any of these parameters can be found in the quality
of the fit apart from only slight differences in the small
deviations from a noise-like distribution of the residuals.

 In contrast, PSR B0833-45 (Vela) shows more complex behavior.
  Only one set of our data can be satisfactorily fitted with a single
  Lorentzian, while a data set taken 8 days earlier and another data set
  taken 7 months later  require a sum of two Lorentzian
  functions with different widths, $w_1$ and $w_2$. 

 It is particularly striking that for the same baseline, MP-HO, with the same
projected length and PA, on 10 May 2012, two Lorentzians are needed for the fit
and on 18 May 2012 only one Lorentzian is needed. Apparently, the scattering
conditions for this pulsar changed drastically over a time span as short as one
week. Anisotropic scattering, indicated by the two Lorentzians changed to
isotropic scattering indicated by the one Lorentzian and then back to
anisotropic scattering.

Earlier, Popov et al. (2019) found already evidence
of anisotropic scattering for this pulsar by comparing the dependence of
visibility amplitude on baseline projection at different baseline position
angles. Our method described in this paper is largely independent of that method
and has advantages with respect to higher signal to noise ratios and more robust
estimates of scattering characteristics.

 That properties of an intervening medium on the line of sight
  to PSR B0833-55 (Vela) differ qualitatively from properties of such
  media for the  other four
  pulsars can  be understood because of peculiarities of PSR B0833-55 (Vela).
  Vela pulsar has the largest mean free electron density along the
  line of sight, the smallest values of $\tscint$ and, together with
  B0329+54, of $\Deltafdif$, and the largest values of $\tausc$ and
  $\thetasc$.  The pulsar is the only one of our sample that is still
  within a visible supernova remnant.

 \citet{popov2019} argued on the basis of the determination of the
  scattering medium position that scintillations of  PSR B0833-55 (Vela) pulsar
  originate at least partly within the supernova remnant. The regions
  responsible for the scattering there differ significantly from the
  standard model of a thin screen.  In particular, the line of
  sight is likely to be nearly tangent to the scattering sheets, which
  are expected to be highly turbulent and rapidly moving. 
  It is therefore conceivable that the scattering screen parameters 
  are highly variable even on such a short time scale of one week.
  In this model, the anisotropic indicatrix and rapid variability of the
  scattering screen parameters are produced naturally.

\section{Conclusions} \label{conc}
 
We present an analysis of two-element interferometry data 
with earth-earth and earth-space baselines 
 for five pulsars: B0329+54, B0823+26, B0834+06, B1933+16 and
B0833-45 (Vela), the latter still embedded in 
its 
 supernova remnant.  The 
cross correlation and autocorrelation functions of the 
interferometer dynamic visibility functions
 in delay 
 and time consist 
of a
band-limited unresolved spike at zero delay 
 lag in $\Delta \tau$ and zero time lag in 
$\Delta t$, interpreted as being due to scintillation noise
(SN), and a smooth slowly varying envelope (SVE).  The
amplitude of the SN spike above the SVE is 
 between 0.6 and 0.95 times 
the amplitude of the SVE which is reminiscent of the
amplitude modulated noise model for pulsar microstructure.  The
amplitude of the SN spike decreases with $\Delta t$ on a time
scale corresponding to the scintillation time, 
 $\tscint$.  The SN for baseline projections
smaller than 
 the size of the diffraction spot, $\rhodif$,
 is uncorrelated with the SN for baseline projections larger than
 $\rhodif$.
 The SVEs in delay lag 
show, supported by numerical
simulations, that they are  well approximated by
one or more Lorentzian functions.  For all pulsars but B0833-45
(Vela), a single Lorentzian only was needed for the fit
indicating isotropic scattering by a thin screen.  For B0833-45
(Vela), mostly at least two Lorentzians with variable widths
were needed for the fit indicating anisotropic scattering
likely in the shell of the supernova remnant and/or the Gum
Nebula with scattering conditions variable on the time scale of
one week or less. 

It is likely that the SVEs of all pulsars can be described by one or more
Lorentzians depending on the complexity of the intermittent scattering material
of the interstellar medium and that fit Lorentzians are a new and more robust
way to describe some scattering properties.

\acknowledgments The RadioAstron project is led by the Astro Space Center of the
Lebedev Physical Institute of the Russian Academy of Sciences and the Lavochkin
Scientific and Production Association under a contract with the Russian Federal
Space Agency, in collaboration with partner organizations in Russia and other
countries. This paper was supported in part by the Russian Academy of Science
Program KP19-270 "The study of the Universe origin and evolution using the
methods of earth-based observations and space research."

\facility{RadioAstron Space Radio Telescope
(Spektr-R), GBT, WSRT, ATCA, Parkes, Ceduna, Mopra, Hobart, 
Hartebeesthoek, Tidbinbilla,  Svetloe
 radio telescope, and Pushchino 22-m radio telescope, used as Tracking Station}.

\software{CFITSIO}

\appendix
\section{Scattering time expressed through the scale parameter of a Lorentzian fit
of a correlation function} \label{tausc-vs-w}

In this appendix we obtain the equation $\tausc=w/2$ used in Sect. \ref{sec:broad}.
The derivation is based on the assumption that the
observed intrapulse variations of fringe visibility magnitude 
can be described by the amplitude-modulated noise (AMN) model developed by
\citet{rickett1975}.  In order to make the relationship with the AMN model
clear, the notations chosen here are close to those used in the cited paper.  In
particular, for a fixed baseline A-B and polarization $P$ ($P=\text{L}$ or $P=\text{R}$) we
designate $I(\tau)=|V^P_{A-B}(\tau)|$.  The functions,
$ACF^{P}_{A-B}(\Delta\tau)$, introduced in Sect. \ref{sec:obs} may be expressed as
$ACF^{P}_{A-B}(\Delta\tau)=\langle R_I(\Delta\tau) \rangle$,
where $\langle\,\rangle$ denotes the ensemble avarage and the operator, $R$,
acting on a random process $z(\tau)$, is defined by
\begin{equation}                                          \label{eq:Rdef}
R_z(\Delta\tau)=\int z(\tau)z(\tau+\Delta\tau)\,d\tau\,
\end{equation}
with integration performed over the total duration of the pulse.
Further, we decompose the observed variability of the visibility magnitude
as $I(\tau)=a^2(\tau)I_1(\tau)$.  Here $a^2(\tau)$ is a deterministic modulating
slowly varying function that reflects the time dependence of $\langle
I(\tau) \rangle$.  The factor, $I_1(\tau)$, is a stationary random process
describing the scintillation noise (SN) that originates in the scattering
matter.  The time scale of SN variations is much smaller than the time scale of
variations of the modulating function.

If we additionaly assume that there exists such a stationary complex Gaussian random
process, $x(\tau)$, that 
\begin{equation}                                          \label{eq:I1_as_xx}
I_1(\tau)\approx |x(\tau)|^2\,,
\end{equation}
than the AMN model is directly applicable to our problem.  The important
consequence of the model is that 
$\langle R_I(\Delta\tau) \rangle =\Rb(\Delta\tau)+\Rc(\Delta\tau)$.
Here, the term,  $\Rb$, is the broad component which varies slowly over the whole range of
$\Delta\tau$, and the term, $\Rc$, represents the narrow central spike with
\begin{eqnarray} 
\Rb(\Delta\tau)& \propto & R_{a^2}(\Delta\tau)\,,     \label{eq:Rb_vs_SVE} \\
\Rc(0)&=&\Rb(0)\,.                                    \label{eq:Rb_vs_Rc}
\end{eqnarray}

The values, $\Rc(0)$, and $\Rb(0)$, can be easily measured observationally.  If
equality (\ref{eq:Rb_vs_Rc}) is satisfied with sufficient precision, then it is likely
that the AMN model is applicable, the equation (\ref{eq:Rb_vs_SVE}) also holds, and
the determination of the modulating function, $a^2(\Delta\tau)$, reduces to the problem
of finding a function with given autocorrelation.

For CCFs
it is difficult, if at all possible, to find a complete analogue to the function,
$I(\tau)$. Thus, the line of reasoning based on direct use of results of
\citet{rickett1975} is not applicable.  But the overall similarity of the
mathematics encountered in considering both the ACFs and CCFs (in all cases we
analize mixed fourth moments of the incident field modulated by a comparatively
slowly varying deterministic function and a stationary random process exhibiting
rapid variations) makes it likely that if equality (\ref{eq:Rb_vs_Rc}) is
satisfied with sufficient precision, then equation (\ref{eq:Rb_vs_SVE}) can be
used to determine the form of the modulating function.

In finding $a^2(\Delta\tau)$ we use for the broad component $\Rb(\Delta\tau)$ of
the measured function $\langle R_I(\Delta\tau) \rangle$ the approximation
\begin{equation}                                          \label{Rb-thru-L}
\Rb(\Delta\tau)=L(\Delta\tau,w,C)\,,  
\end{equation}
where $L(\Delta\tau,w,C)= rw/(\Delta\tau^2+w^2)+C$.  We assume that the constant
offset, $C$, is caused only by errors in the determination of baselines of the
individual visibility magnitudes, that is $C=0$ in equation (\ref{Rb-thru-L}).
Using the identity $R_{L(\Delta\tau,y,0)}=rp L(\Delta\tau,2y,0)$, where the operator,
$R$, is defined in equation (\ref{eq:Rdef}), we obtain from
equation (\ref{eq:Rb_vs_SVE}) that $\tausc=w/2$, where the scattering time, $\tausc$,
is defined as the HWHM of $a^2(\Delta\tau)$.
\bibliographystyle{aasjournal}
\bibliography{Substructure_of_visibility}
\end{document}